# EINSTEIN WAS WRONG: NEWTONIAN DYNAMICS CAN DISAGREE COMPLETELY WITH RELATIVISTIC DYNAMICS AT LOW SPEED

Boon Leong Lan
School of Engineering and Science, Monash University, 2 Jalan Kolej,
46150 Petaling Jaya, Selangor, Malaysia

ABSTRACT

According to Einstein, the trajectory of a particle that is predicted by special relativistic mechanics is well approximated by the trajectory predicted by Newtonian mechanics if the particle speed is low, i.e., much less than the speed of light. However, in this paper, it is shown with a counterexample Hamiltonian dynamical system that Newtonian dynamics can eventually disagree completely with relativistic dynamics even though the particle speed is low. After the breakdown of the agreement, either the Newtonian description is no longer correct or the relativistic description is no longer correct or both descriptions are no longer correct for the low speed motion.

According to Einstein (*1*):

> 'Classical mechanics required to be modified before it could come into line with the demands of the special theory of relativity. For the main part, however, this modification affects only the laws for rapid motions, in which the velocities of matter *v* are not very small as compared with the velocity of light. … **for other motions** [*v* very small compared with the velocity of light] **the variations from the laws of classical mechanics are too small to make themselves evident in practice**.'

In other words, according to Einstein (*1,2*), and also others (*3-6*) later, if the speed of a particle $v$ is much less than the speed of light $c$, then the dynamical prediction of special relativistic mechanics is well approximated by the prediction of Newtonian mechanics. However, in this paper, I will show with a counterexample dynamical system that Einstein was wrong, i.e., Newtonian dynamics can eventually deviate completely from relativistic dynamics even though the particle speed is low, i.e., $v \ll c$.

The counterexample is a one-dimensional Hamiltonian, i.e., nondissipative, dynamical system in a sinusoidal potential that is periodically turned on only for an instant. This periodically delta-kicked system could either be a pendulum in a time-varying gravitational field (*7,8*) or an electron in a time-varying electric field in a plasma (*9,10*).

Newton's equation of motion for such a kicked system is easily integrated exactly (*7,8*) to produce a mapping, called the standard map, of the dimensionless scaled position *X* and dimensionless scaled momentum *P* from $(X_{n-1}, P_{n-1})$, the values just before the *n*th kick, to $(X_n, P_n)$, the values just before the (*n+1*)th kick:

$$P_n = P_{n-1} - \frac{K}{2\pi}\sin(2\pi X_{n-1}) \qquad 1(a)$$

$$X_n = (X_{n-1} + P_n) \bmod 1 \qquad 1(b)$$

where $n = 1, 2, \ldots$, and *K* is a dimensionless positive parameter. The map above has also served as an important model in the field of nonlinear dynamics because a number of dynamical problems are approximately reduced to it (*7,11*).

The relativistic equation of motion is also easily integrated exactly (*9,10*) to produce a mapping of the dimensionless scaled position *X* and dimensionless scaled momentum $P$ from $(X_{n-1}, P_{n-1})$, the values just before the $n$th kick, to $(X_n, P_n)$, the values just before the $(n+1)$th kick:

$$P_n = P_{n-1} - \frac{K}{2\pi}\sin(2\pi X_{n-1}) \qquad \text{2(a)}$$

$$X_n = \left(X_{n-1} + \frac{P_n}{\sqrt{1+\beta^2 P_n^2}}\right) \bmod 1 \qquad \text{2(b)}$$

where $n = 1, 2, ...$, and in addition to $K$, there is another dimensionless positive parameter $\beta$.

Detailed properties of the Newtonian standard map [Eqs. 1(a) and 1(b)] and the relativistic standard map [Eqs. 2(a) and 2(b)] can be found in references (*7,11*) and (*9,10,12*) respectively.

In general, according to special relativistic mechanics,

$$\frac{p}{m_0 c} = \frac{\frac{v}{c}}{\sqrt{1-\left(\frac{v}{c}\right)^2}} \qquad (3)$$

where $p$ is the relativistic momentum and $m_0$ is the rest mass. Eq. (3) implies that

$$\frac{v}{c} = \frac{\frac{p}{m_0 c}}{\sqrt{1+\left(\frac{p}{m_0 c}\right)^2}}. \qquad (4)$$

For the relativistic standard map [Eqs. 2(a) and 2(b)], since $\beta P$ ($P$ is the dimensionless scaled momentum) is (*10*) equal to $\frac{p}{m_0 c}$, we can easily see from Eq. (4) that $\beta P \ll 1$ implies $v \ll c$. So, conventionally, the phase space trajectory generated by the Newtonian standard map is expected to agree with the phase space trajectory generated by the relativistic standard map for $\beta P \ll 1$ since $v \ll c$. This expectation seems reasonable at first sight because the Newtonian standard map [Eqs. 1(a) and 1(b)], which differs from the relativistic standard map only by the square-root term in Eq. 2(b), 'approximates' the relativistic standard map if $\beta P \ll 1$. However, in all the cases that have been studied, the Newtonian trajectory eventually deviates completely from the relativistic trajectory although $v \ll c$. An example of this counter-expected behavior is given next.

In this example, the dimensionless scaled momentum $P$ of the relativistic trajectory, with initial conditions $X_0 = 0.5$ and $P_0 = 99.9$, generated by the relativistic standard map, with parameters $K = 0.9$ and $\beta = 10^{-7}$, is always $\approx 100$ (in particular, $P$ is always bounded between 99.6 and 100.4). Therefore, $\beta P \approx 10^{-5}$, which implies, according to Eq. (4), that $v \approx 10^{-5} c$, or in other words, $v$ is always merely 0.001% of $c$ (for comparison, the orbital speed of the earth is about 10 times faster). At such low speeds (*4*), it is conventionally expected that the Newtonian trajectory, which is generated by the Newtonian standard map with the same parameter $K$ and initial conditions, would agree with the relativistic one.

To compare the two trajectories above, the Newtonian and relativistic positions are plotted versus iteration in Fig. 1 while the Newtonian and relativistic momentum are plotted versus iteration in Fig. 2 for the first 140 iterations. In each figure, successive values of the dynamical quantity predicted by each theory are connected by straight line to aid the eyes. The following behavior for both the position and momentum can clearly be seen in the two figures. The Newtonian values agree with the relativistic values for the first 113 iterations. However, contrary to conventional expectation, the Newtonian values no longer agree with the relativistic ones from iteration 114 onwards. The plotted Newtonian and relativistic values, which were computed in double precision (15 significant figures), are converged, where the degree of convergence decreases with increasing number of iterations. The degree of convergence was established using the standard method (*11*) of varying the numerical precision: here, by comparing the double precision values to the corresponding values computed in long double precision (18 significant figures). Hence, the breakdown of the agreement of the Newtonian values with the relativistic values, for both position and momentum, after iteration 113 is not a numerical artifact.

In all the other cases (different parameters, different initial conditions, different very-small values of $v/c$), the convergence check also shows that the breakdown of the agreement of the Newtonian dynamics with the relativistic dynamics at low speed is not a numerical artifact.

In summary, I have shown with a counterexample Hamiltonian dynamical system that, contrary to Einstein's claim, the agreement of Newtonian dynamics with relativistic dynamics can eventually break down even though the particle speed is low, i.e., much less than the speed of light. After the breakdown of the agreement, either the Newtonian description is no longer correct or the relativistic description is no longer correct or both descriptions are no longer correct for the low speed motion.

I thank Yong Shaohen for coding the mappings in C for the convergence checks.

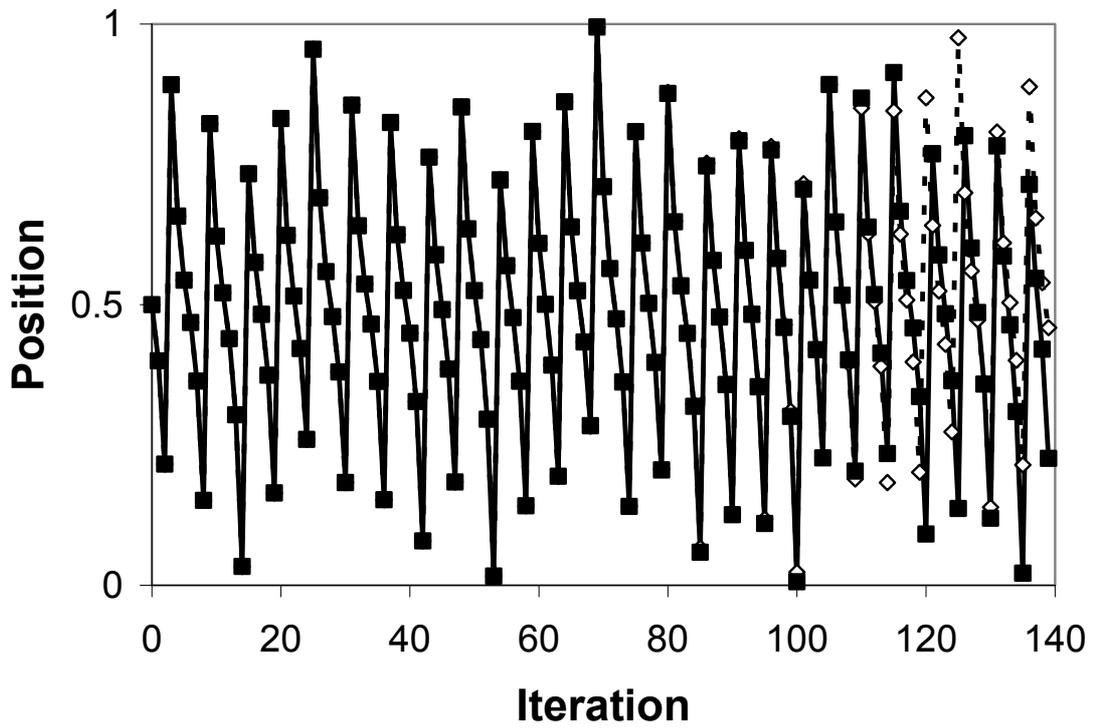

FIGURE 1    Comparison of the Newtonian Position (diamonds) with the Relativistic Position (squares)

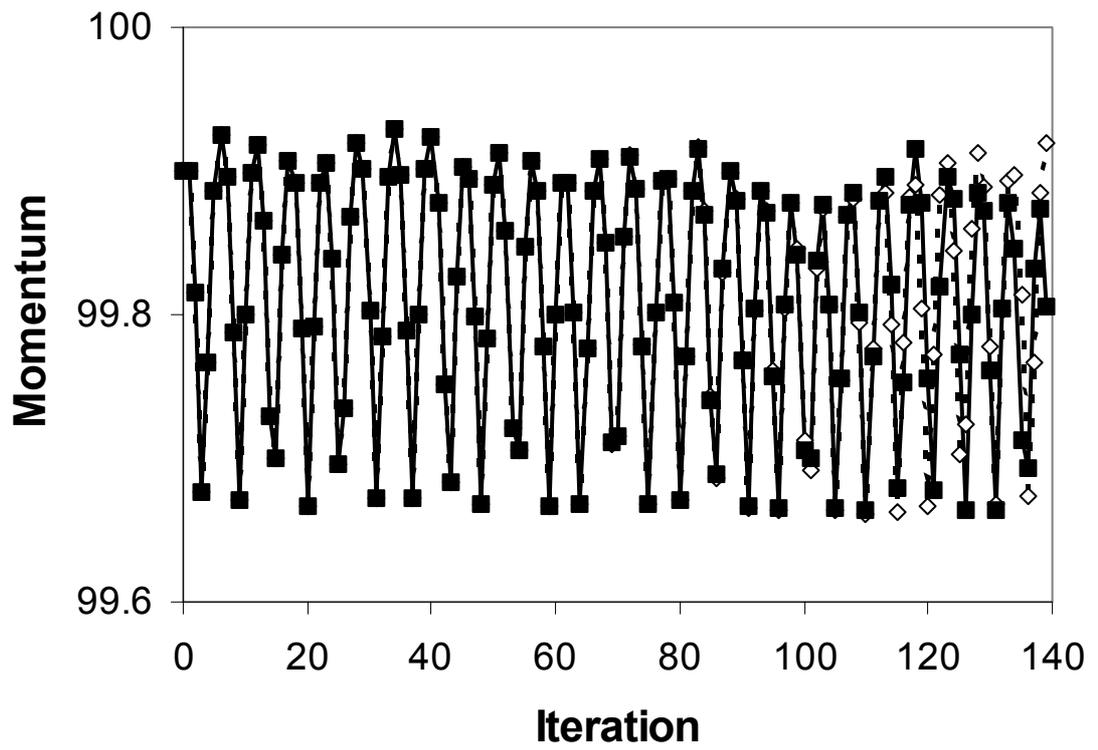

FIGURE 2   Comparison of the Newtonian Momentum (diamonds) with the Relativistic Momentum (squares)